\documentclass[pra,final,twocolumn]{revtex4-2}
\usepackage{amsmath}
\usepackage{graphicx}
\usepackage{amsfonts}
\usepackage{amssymb}
\usepackage{subfigure}
\usepackage{lineno}

\begin{document}

\title{Decoherence, Entanglement, and Information in the Electron Double-Slit Experiment with Monitoring }

\author{Frederick W. Strauch}
\affiliation{ 
Department of Physics, Williams College, Williamstown, MA 01267
}%

\date{\today}

\begin{abstract}
This paper considers a theoretical model of the double-slit experiment with electrons whose paths are monitored.  This monitoring, inspired by a recent text by Maudlin, is performed by the Coulomb scattering of the electron by a proton.  A simple quantum mechanical calculation is presented, inspired in part by a recent experimental demonstration of this famous thought experiment.  The results illustrate the relationship between entanglement and the loss of coherence in the interference pattern.  The tradeoff between the visibility of interference and the information gained by measurement is also explored.  This calculation can provide advanced undergraduates insight into decoherence, entanglement, and quantum information. 
\end{abstract}

\maketitle

\section{\label{sec:Introduction}Introduction}

The double-slit experiment is the classic thought experiment in quantum mechanics.  Famously, Feynman characterized this ``experiment with the two holes''  as containing  ``all of the mystery of quantum mechanics''  \cite{[{}][{, Chapter 6.  A video of the Messenger lectures can be found at https://www.feynmanlectures.caltech.edu/messenger.html}]feynman2017character}.  Modern realizations of this experiment have demonstrated single-electron detection \cite{frabboni2012young} and {\em in situ} control of the slits \cite{bach2013controlled}.  These complement the ever expanding range of diffraction experiments with single neutrons \cite{zeilinger1988single,zeilinger1999experiment}, atoms \cite{cronin2009optics}, and molecules \cite{nairz2003quantum,brand2021single}, which have demonstrated matter wave interference at ever larger scales.  

The quantum mystery (or at least one such mystery) arises when one considers how a single particle is able to exhibit interference.  This seems to require the particle to pass through both slits at once.  Consider the arrangement shown in Fig.~1, where an electron passes through a double-slit (at $t=0$).  Should one ensure that the particle passes through slit 1 (by closing slit 2), the probability distribution for measuring the particle at some later time and at some distance $x$ is given by $|\psi_1(x)|^2$, where $\psi_1(x)$ is the wave function for the particle at time $t$ (note that, for simplicity, the dependence of the wave function on the distance $z$ has been suppressed \footnote{\label{znote}In this paper the time-dependent wave function of the electron is treated as function of $x$ only.  One could extend this model by using electron wave packets $\psi(x,z,t)$ whose peaks move along the trajectory $z = v t$, where $v$ is the mean velocity.  This would not change the main results of this paper.}).  Similarly, for a particle sure to pass through slit 2 with a corresponding wave function $\psi_2(x)$, the probability distribution is given by $|\psi_2(x)|^2$.  However, should both slits be open and the particle described by the (unnormalized) superposition $\psi(x) = \psi_1(x) + \psi_2(x)$, the resulting probability distribution is not 
\begin{equation}
P_{\text{1\,or\,2}}(x) = |\psi_1(x)|^2 + |\psi_2(x)|^2,
\end{equation}
corresponding to the particle passing through slit 1 or slit 2, but rather 
\begin{equation}
P_{\text{1\,and\,2}}(x) = |\psi_1(x) + \psi_2(x)|^2,
\end{equation}
which exhibits a classic interference pattern (shown at the far right in Fig.~1).  And yet, when the slits are monitored, the interference is destroyed, and the probability distribution ``collapses'' from $P_{\text{1\,and\,2}}(x)$ to $P_{\text{1\,or\,2}}(x)$. This transition from a quantum superposition to a classical probability is, indeed, a persistent mystery.

\begin{figure}[ht]
\includegraphics[width = 3.25in]{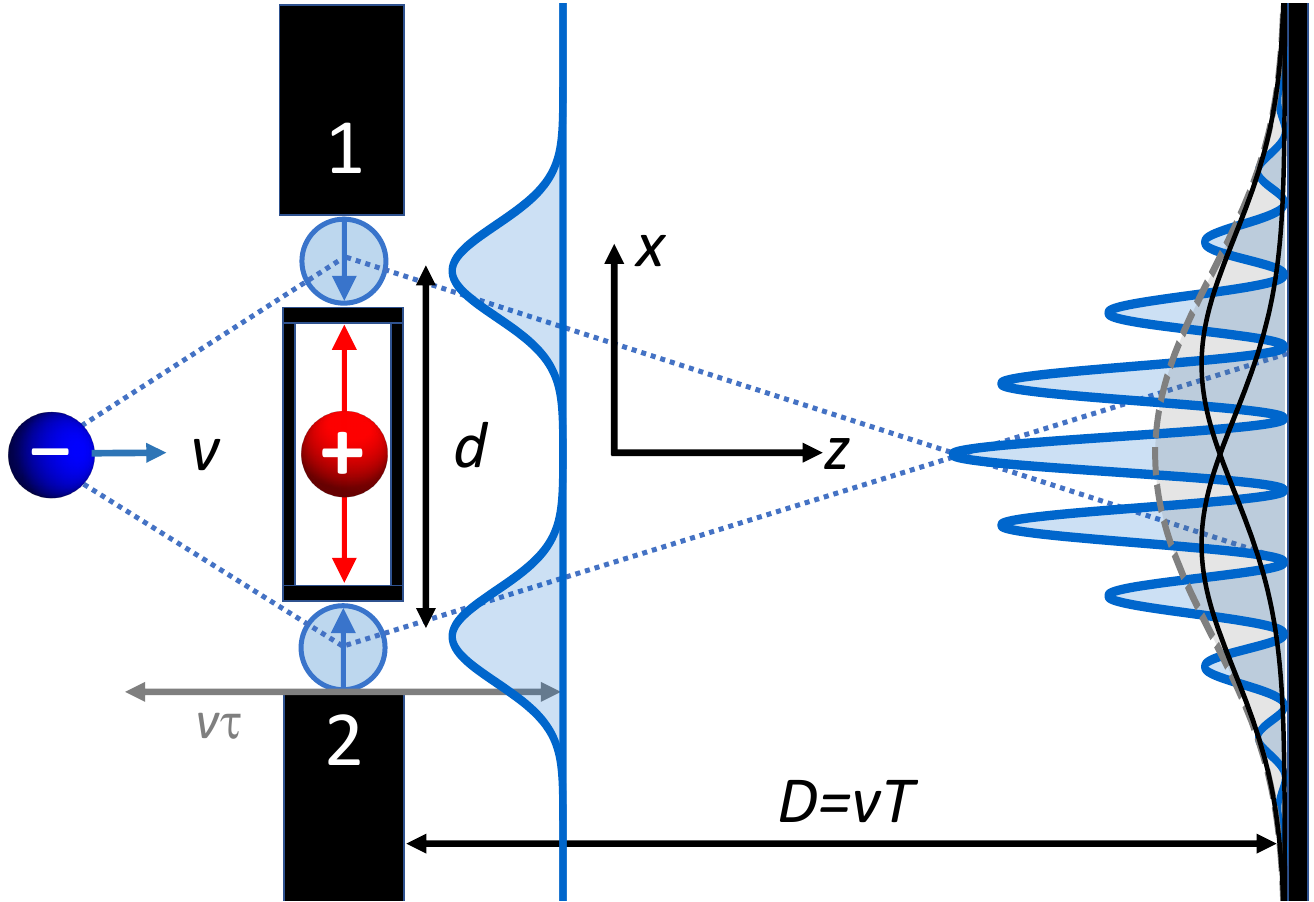}
\caption{\label{fig1} Schematic of the double-slit experiment with monitoring.  An electron (in blue) travels to the right ($z$ direction) with velocity $v$.  Its wave function, as a function of position $x$ (perpendicular to its velocity) is initially a superposition of states localized about slits 1 and 2; the slits are separated by a distance $d$.  Between the slits sits a proton (in red), which interacts with the electron for times between $-\tau/2$ and $\tau/2$.  Depending on the momentum exchanged between the electron and the proton (vertical arrows), the probability, when observed at a distance $z=D=v T$ from the slits (on the right) will exhibit an interference pattern ranging from full (blue, solid) to zero (gray, dashed) coherence, corresponding to the probability distributions $P_{\text{1\,and\,2}}(x) = |\psi_1(x) + \psi_2(x)|^2$ and $P_{\text{1\,or\,2}}(x) = |\psi_1(x)|^2 + |\psi_2(x)|^2$, respectively (note that, for simplicity, the dependence of the wave function on $z$ has been suppressed).  The single-slit probability distributions $|\psi_1(x)|^2$ and $|\psi_2(x)|^2$ are modeled by spreading Gaussian wave packets and are indicated for times $t \approx \tau/2$ (blue, left) and $t = T = D/v$ (black, right).}
\end{figure}

Explaining this mystery has a long and distinguished history.  In an early round of the Bohr-Einstein debates, Einstein devised a challenge to quantum theory by introducing a movable screen with a slit that would allow one to measure the path of a passing particle.  For a double-slit experiment, this would allow one to deduce through which slit the particle passes.  In response, Bohr argued \cite{Bohr1949}, using Heisenberg's uncertainty principle---applied to both the particle and the screen---that such a measurement would introduce some uncertainty in the positions of the interference fringes.  This uncertainty excludes observation of a clear interference pattern.  This is in accord with the principle of complementarity, here for wave-particle duality: a measurement of the particle's path is complementary to the observation of its wave interference.  

A somewhat quantitative account of this loss of interference was supplied in the text by Bohm \cite{bohm2012quantum}.  He argued that the interaction of the particle with a macroscopic apparatus that monitors its position will modify the particle's wave function by
\begin{equation}
\psi(x) \to  \psi_1(x) e^{i \varphi_1}+ \psi_2(x) e^{i \varphi_2},
\label{Bohmphase}
\end{equation}
where the phases $\varphi_1$ and $\varphi_2$ are effectively random variables \footnote{A qualitative argument is given in Ch. 6 Secs. 3 and 4 of Ref. 11.  The phases arise from an interaction between the system and the measuring apparatus.   A more careful analysis is found in Secs. 5-10 of Ch. 22, using the example of a Stern-Gerlach measurement of an electron's spin.   Bohm's analysis identifies what we now call entanglement: ``there is clearly no single wave function belonging to the spin alone, but, instead, there is only a combined wave function in which spin and apparatus co-ordinates are inextricably bound up'' (p. 603 of Ref. 11)}.  Thus, while the probability distribution for any individual outcome will be given by
\begin{align}
P_{\varphi}(x) = & |\psi_1(x)|^2 + |\psi_2(x)|^2  \nonumber \\
& +  e^{-i \varphi} \psi_1^*(x) \psi_2(x)  + e^{i \varphi} \psi_2^*(x) \psi_1(x),
\end{align}
the total probability will be a statistical average over the random phase $\varphi = \varphi_1 - \varphi_2$.  This process is known as dephasing, and is a simple form of decoherence ({\em i.e.} the loss of coherence) \cite{schlosshauer2007decoherence}.  The resulting probability distribution is
\begin{align}
\langle P_{\varphi}(x) \rangle =  & |\psi_1(x)|^2  + |\psi_2(x)|^2 \nonumber \\
 & + \mathcal{V} \left[ \psi_1^*(x) \psi_2(x)  + \psi_2^*(x) \psi_1(x) \right],
 \label{partialP}
\end{align}
where the fringe visibility $\mathcal{V} = \langle e^{-i \varphi} \rangle \le 1$ is taken to be real.  For complete dephasing, $\mathcal{V} \to 0$, and $\langle P_{\varphi}(x) \rangle  \to P_{\text{1\,or\,2}}(x)$.  

An even more quantitative account was developed by Wootters and Zurek \cite{wootters1979complementarity}.  Central to Bohr's response to Einstein (and important for later rounds of their debate) was the need to consider the quantum interaction between the particle and the screen.  Wootters and Zurek supplied just such an account of this interaction, appropriate for the double-slit experiment with photons, by treating the screen as a quantum harmonic oscillator and modeling the interaction by an exchange of momentum between the passing photon and the movable screen.  The resulting probability distribution agrees with Eq.~(\ref{partialP}) with a visibility that depends on this exchange of momentum (a similar calculation, for a neutron interferometer, can be found in \cite{greenberger1983neutron}).  They also established an entropy inequality to quantify the wave-particle duality of the photon, in terms of how much information one can obtain about the particle's position (the so-called ``which-way'' information) while maintaining a finite visibility for the interference fringes.  Subsequent work has extended this duality to more general interferometric settings \cite{greenberger1988simultaneous, jaeger1993complementarity, jaeger1995two,englert1996fringe}, and this remains a topic of great interest \cite{jakob2010quantitative,qian2018entanglement, qian2020turning,maleki2023revisiting}.   An introduction to quantitative wave-particle duality was given by Qureshi \cite{qureshi2016quantitative}. 

Wootters and Zurek further argued that the loss of interference is due to the entanglement between the particle and the screen.  This is a more general source of decoherence than dephasing \footnote{Note that Schlosshauer does not consider all dephasing processes to be examples of ``true'' decoherence (see Sec. 2.12 in Ref. 13).  While dephasing is a common mechanism for the loss of coherence, decoherence due to entanglement is the subject of this paper}, and entanglement-induced decoherence is now considered to be the process by which quantum probabilities become effectively classical \cite{zurek2003decoherence,schlosshauer2019quantum}.  For the double-slit experiment, the monitoring of the particle's position by some detector \cite{qureshi2016quantitative} leads to states of the form
\begin{equation}
\frac{1}{\sqrt{2}} |\psi_1\rangle |d_1\rangle + \frac{1}{\sqrt{2}} |\psi_2\rangle |d_2\rangle
\end{equation} 
where I have used Dirac notation and introduced the detector states $|d_1\rangle$ and $|d_2\rangle$.  As will be explained below, when the detector states become orthogonal, the interference pattern goes to zero.  In fact, the visibility $\mathcal{V} = \langle d_1 | d_2\rangle$ corresponds to the off-diagonal elements of the particle's density matrix, whose decay is traditionally associated with decoherence. This decay can be due to uncontrolled interaction and subsequent entanglement of the system with the environment, which acts as a detector.  A pedagogical treatment of environment-induced decoherence for the double-slit with monitoring can be found in Sec. 2.6 of the text by Schlosshauer \cite{schlosshauer2007decoherence}.  A more detailed calculation of measurement-induced decoherence for the double-slit was given by Kincaid, McLelland, and Zwolak \cite{kincaid2016measurement}.  For both cases, when there is only partial entanglement between the system and the detector, the resulting probability distribution exhibits the same partial interference seen in Eq.~(\ref{partialP}).  For both environment- and measurement-induced decoherence maximal entanglement, corresponding to perfect monitoring of the particle, yields $\mathcal{V}=0$.  

In this paper, I provide a quantitative account of another form of the double-slit experiment with monitoring, appropriate for electrons, that was recently described by Maudlin 
\cite{[{}][{.  The double-slit with monitoring appears on pp. 14-17, 53-59, and 145-156.}]maudlin2019philosophy}.  This model, illustrated in Fig.~1, involves an interaction between the electron and another charged particle, such as a proton.  Much like the Bohr-Einstein scenario, momentum is transferred between the passing electron and the proton.  Much like the Wootters-Zurek analysis, it is the entanglement between the electron and the proton that leads to the loss of coherence.  However, the interaction here is simply the Coulomb force between the proton and the electron.  Maudlin presented this model to provide a simple and (at least qualitatively \footnote{\label{mnote}Figures 16 (on p. 57), 25 (on p. 147) and 26 (on p. 156) in Maudlin (Ref. 29) attribute the loss of coherence to the displacement of the proton in the electron-proton configuration space.  The model developed here shows that decoherence requires only a separation of the proton's states in momentum space, and not a subsequent displacement in position space.}) visualizable example of decoherence.  This particular form of entanglement-induced decoherence is not purely theoretical, having been observed in the photoionization of $\text{H}_2$, in which the two protons acted as the slits and one electron (excited at lower energy) acted as the environment for the other (higher energy) electron \cite{akoury2007}.  This paper aims to provide a simplified calculation of entanglement-induced decoherence of the double-slit experiment with electrons that could reasonably be included in an undergraduate course in quantum mechanics.  

In addition to this calculation, I provide a detailed accounting of the information that can be gained about the position of the electron by measurements on the proton, and how this information depends on the type of measurements performed and the degree of entanglement.  This provides a simple physical example to illustrate a number of important topics from the field of quantum information \cite{nielsen2010quantum,barnett2009quantumbook}.  Altogether, this paper provides a pedagogical framework for studying decoherence, entanglement, and information in this classic experiment.

This paper is organized as follows. The basic model is presented in Sec.~II, in which the exchange of momentum between the electron by the proton is treated in a semiclassical fashion.  Using Gaussian wave functions for the electron and the proton, the fringe visibility $\mathcal{V}$ is calculated and the relevant parameters are estimated.  Since these wave functions are Gaussian, the time evolution for the interference pattern can be calculated exactly.  The pattern is explored in Sec.~III using the experimental parameters of Bach {\it et al.} \cite{bach2013controlled}.  The entanglement between the electron and the proton occurring in this model, as a function of the visibility $\mathcal{V}$, is calculated in Sec.~IV.  The relationship between the visibility of the interference pattern and information that can be gained about the electron's position is explored in Sec.~V, using the measurement methods discussed by Maudlin \cite{maudlin2019philosophy},  Wootters and Zurek  \cite{wootters1979complementarity}, and Qureshi \cite{qureshi2016quantitative}.  Finally, I conclude in Sec.~VI with a discussion of the possible use of this model to discuss the fundamentals of decoherence, entanglement, and information.  Explicit details for the calculation of the interference pattern and the information gains are found in the Appendices.



\section{Coulomb Scattering Model}

The basic setup for the double-slit experiment with monitoring is shown in Fig.~1.  The two slits are separated by distance $d$ (along the $x$-direction), and the electron passes through them (along the $z$-direction) with (mean) velocity $v$.  Following Maudlin \cite{maudlin2019philosophy}, I introduce a small chamber between the two slits, in which sits a proton.  The proton is free to move toward one or the other slit, but is otherwise contained in the chamber.  As the electron passes through slit 1 or slit 2, the Coulomb force on the proton will cause it to move ``up'' or ``down.''  This section develops a quantum mechanical model of this interaction.  An initial wave function for the electron and proton, each restricted to one spatial dimension (along the $x$-direction, see $[9]$) is specified and a simple model is used to calculate the effect of the interaction on this wave function (a more detailed calculation is discussed in Appendix A).  The resulting interference pattern is found by propagating this wave function from the slits to the screen, and will be presented in Sec.~III.

 The initial wave function for the electron, just before interacting with the proton, is chosen to be the superposition
\begin{equation}
\psi_{\text{e}}(x) = \frac{1}{\sqrt{2}} \psi_1(x) + \frac{1}{\sqrt{2}} \psi_2(x),
\end{equation}
where each wave function is of Gaussian form, {\em i.e.}
\begin{equation}
\psi_1(x) = (\pi \delta^2)^{-1/4} \exp \left[-\frac{1}{2} \frac{(x-d/2)^2}{\delta^2} \right]
\label{psi1}
\end{equation}
and
\begin{equation}
\psi_2(x) = (\pi \delta^2)^{-1/4} \exp \left[ -\frac{1}{2} \frac{(x+d/2)^2}{\delta^2} \right].
\label{psi2}
\end{equation}
This choice allows for an exact calculation of the time-evolution of the interference pattern.  Note that I have neglected the overlap between $\psi_1(x)$ and $\psi_2(x)$ in the normalization for $\psi_{\text{e}}(x)$, assuming that $\delta \ll d$; the proper normalization will be given in Sec.~III.  The proton's wave function is also taken to be Gaussian:
\begin{equation}
\psi_{\text{p}}(X) = (\pi \Delta^2)^{-1/4} \exp \left( - \frac{1}{2} \frac{X^2}{\Delta^2} \right),
\label{psip}
\end{equation}
where $X$ denotes the position of the proton.  The total wave function going into the double-slit is thus
\begin{equation}
\Psi_{\text{in}}(x,X) = \frac{1}{\sqrt{2}} \psi_1(x) \psi_{\text{p}}(X) + \frac{1}{\sqrt{2}} \psi_2(x) \psi_{\text{p}}(X).
\label{psiin}
\end{equation}

When the electron passes through the double-slit, it is scattered by the Coulomb interaction with the proton.  A simple model of this interaction uses the momentum exchange found from by the classical force between the electron and the proton. Specifically, for an electron passing through slit $1$, the Coulomb force on the electron (along the $x$-direction, see Fig.~1) is calculated as
\begin{equation}
F(t) = - \frac{e^2}{4 \pi \varepsilon_0} \frac{d/2}{\left[ d^2/4 + (v t)^2 \right]^{3/2}}.
\end{equation}
Here I have set $z = v t$ and used the average value of $|x-X| = d/2$, assuming that the uncertainty ($\sqrt{\delta^2 + \Delta^2}$) is much less than $d$.  The scattering of the electron and the proton will involve the total impulse
\begin{align}
P = \int |F(t)| dt &= \frac{e^2}{4 \pi \varepsilon_0} (d/2) \int_{-\tau/2}^{\tau/2} \frac{dt}{\left[ d^2/4 + (v t)^2 \right]^{3/2}} \nonumber \\
& = \frac{e^2}{4 \pi \varepsilon_0} \left( \frac{4}{v d} \right) \frac{(v \tau / d)}{\sqrt{1 + (v \tau/d)^2}} \nonumber \\
& \approx \frac{e^2}{4 \pi \varepsilon_0} \left( \frac{4}{v d} \right),
\label{scattering}
\end{align}
where I have introduced the interaction time $\tau$ and used the approximation $\tau \gg d/v$ (the momentum reaches $99.5\%$ of its final value for $\tau = 10 d/v$); note that the impulse along the $z$-direction equals zero.  This semiclassical model is appropriate provided the scattering is small ($P \ll m v$) and the interaction time is small compared to the characteristic spreading time ($\tau \ll m \delta^2 / \hbar$), so that there is little spreading of the electron's wave packet during its interaction with the proton.  This will be checked below in Sec.~III.  This result for $P/(m v)$ agrees with the small-angle approximation of the Rutherford scattering angle with impact parameter $d/2$.   

Having determined the exchange of momentum, the total wave function going out of the double-slit is given by the following: 
\begin{align}
\Psi_{\text{out}}(x,X) = & \frac{1}{\sqrt{2}} \psi_1(x) e^{-i P x/\hbar} \psi_{\text{p}}(X) e^{+i P X/\hbar}  \nonumber \\
& + \frac{1}{\sqrt{2}} \psi_2(x) e^{+i P x/\hbar} \psi_{\text{p}}(X) e^{-i P X /\hbar}.
\label{psiout}
\end{align}
Here the components of the wave function have been multiplied by the appropriate momentum shifts $e^{\mp i P x/\hbar} e^{\pm i P X/\hbar}$, where $P$ is the momentum calculated in Eq.~(\ref{scattering}).  The probability distribution for the electron can be calculated by integrating the absolute square of this wave function over the position of the proton:
\begin{equation}
P_{\text{e}}(x) = \int_{-\infty}^{\infty} |\Psi_{\text{out}}(x,X)|^2 dX.
\end{equation}

At this point, it is interesting to compare Eq.~(\ref{psiout}) to the wave function of Eq.~(\ref{Bohmphase}) for Bohm's dephasing argument.  The momentum exchange has changed the electron states $\psi_1(x) \to \psi_1(x) e^{-i P x/\hbar}$ and $\psi_2(x) \to \psi_2(x) e^{+i P x/\hbar}$, while also imparting the phases $\varphi_1 = P X/\hbar$ and $\varphi_2 = - P X/\hbar$.  Due to the uncertainty in the proton's position, these phases are effectively random, and thus the visibility can be calculated as 
\begin{equation}
\mathcal{V} = \langle e^{-i \varphi} \rangle = \int_{-\infty}^{\infty} e^{-i 2 P X/\hbar} |\psi_{\text{p}}(X)|^2 dX = e^{- P^2 \Delta^2 /\hbar^2}.
 \label{visibility1}
\end{equation}
As discussed above, the reduction in visibility is a measure of decoherence.  

It is also interesting to re-express the proton wave functions $\psi_p(X) e^{\pm i P X/\hbar}$ in momentum space by their Fourier transform.  These can be written as
\begin{equation}
\Phi_1(k) = \pi^{-1/4}  \Delta^{1/2} \exp \left[ -\frac{1}{2} \Delta^2 (k - P/\hbar)^2 \right].
\label{momentum1}
\end{equation}
and 
\begin{equation}
\Phi_{2}(k) = \pi^{-1/4}  \Delta^{1/2} \exp \left[ -\frac{1}{2} \Delta^2 (k + P/\hbar)^2 \right],
\label{momentum2}
\end{equation}
with normalization $\int |\Phi_1|^2 dk = \int |\Phi_2|^2 dk = 1$.  These two states correspond to the proton moving ``up'' and ``down'', respectively, with average momentum $\pm P$.  However, these states are not orthogonal.  Their overlap is precisely the visibility of Eq.~(\ref{visibility1}), {\em i.e.}
\begin{equation}
\mathcal{V} = \int_{-\infty}^{+\infty} \Phi_1^*(k) \Phi_2(k) dk =  e^{- P^2 \Delta^2 /\hbar^2}.
\end{equation}
From this perspective, the visibility is associated with how ``close'' the states $\Phi_1(k)$ and $\Phi_2(k)$ are to each other.  This, in turn, is associated with how correlated the proton is with the position of the electron.  This will be discussed further in Sec.~IV and Appendix B.

\section{Interference Pattern}

The interference pattern is found by solving the time-dependent Schr{\"o}dinger equation for $\Psi(x,X,t)$ given the initial condition $\Psi(x,X,t=0) = \Psi_{\text{out}}(x,X)$.  This can be found analytically by using the time evolution of free Gaussian wave packets (here in an entangled superposition), and is detailed in Appendix A.  The resulting probability distribution is given by
\begin{align}
P_{\text{e}} (x,t) & = \int_{-\infty}^{+\infty} |\Psi(x,X,t)|^2 dX \nonumber \\
& = \frac{N^2}{2} \sqrt{A/\pi} \left[ e^{-A (x+v_0 t-d/2)^2} + e^{-A(x-v_0 t+d/2)^2} \right. \nonumber \\
& \quad \quad \quad \left. +  2 \mathcal{V} e^{-A [x^2 +(d/2 - v_0 t)^2]} \cos(2 \pi x/\Lambda) \right],
\label{patternp}
\end{align}
where
\begin{align}
A &= \frac{\delta^2}{\delta^4 + (\hbar t/m)^2} \approx \left( \frac{m \delta/t}{\hbar} \right)^2, \nonumber \\
v_0 &= P/m = \alpha \hbar / (m d) \approx \frac{e^2}{4 \pi \varepsilon_0} \left( \frac{4}{m v d} \right), \nonumber \\
\mathcal{V} &=  e^{- P^2 \Delta^2 /\hbar^2} = e^{- \alpha^2 \Delta^2/d^2 }, \nonumber \\
\frac{\Lambda}{2\pi} &= \frac{(\hbar t/m)^2 + \delta^4}{d (\hbar t/m) + 2 P \delta^4/\hbar} \approx \frac{\hbar}{m d/t}.
\label{patterneq}
\end{align}
Here I have introduced the wave packet spreading factor $A$, the recoil velocity $v_0$, the fringe spacing $\Lambda$, and the dimensionless interaction parameter
\begin{equation}
\alpha = P d / \hbar \approx \frac{e^2}{\pi \varepsilon_0 \hbar v}. 
\label{visibility2}
\end{equation}
The approximations for $A$ and $\Lambda$ are for times larger than the characteristic spreading time ($t \gg m \delta^2 / \hbar$).  Finally, the normalization constant is given by
\begin{equation}
N = \left( 1 + e^{-(d/2)^2/\delta^2}e^{ -P^2 \delta^2/\hbar^2}e^{ -P^2 \Delta^2/\hbar^2} \right)^{-1/2}.
\end{equation}
which is approximately $1$ for $\delta \ll d$.  

To illustrate how this model could, in principle, correspond to a physical experiment, I consider interference patterns corresponding to those observed by Bach {\it et al.} \cite{bach2013controlled}.  This should not be considered as an accurate model of their experiment, but rather as a pedagogical tool to introduce the basic ideas of decoherence to undergraduates in the context of an elegant realization of a classic thought experiment.  Identifying more realistic decoherence models for actual electron interference experiments is a topic of great interest \cite{howie2011mechanisms, beierle2018experimental, chen2020dephasing, kerker2020quantum, chen2021absence}.

To estimate $\alpha$, we need the velocity of electrons.  Bach {\it et al.} used $600 \ \mbox{eV}$ electrons, corresponding to a velocity $v = 1.45 \times 10^{7} \ \mbox{m/s}$, which when substituted into Eq.~(\ref{visibility2}) yields $\alpha = 0.6$.  Note that faster electrons would scatter less with smaller values of $\alpha$.  Conversely, slower electrons would have correspondingly larger values.  For the remaining parameters, the double-slit had a slit separation of $d = 272 \ \mbox{nm}$ and was observed at a distance $D = 240 \ \mbox{mm}$, corresponding to a propagation time $T = D/v = 16.5 \ \mbox{ns}$.  The experimental results (Fig.~S2 in their Supplementary Information), are reproduced by using a width parameter for the initial Gaussian (for the electron) of $\delta = 20 \ \mbox{nm}$.   Note that $T$ is much larger than the characteristic spreading time $m \delta^2 /\hbar = 3.45\ \mbox{ps}$.  The resulting interference pattern, with $\mathcal{V} = 0.8$, appropriate for the experiment of Bach {\it et al.} \cite{bach2013controlled} is shown in Fig.~2(a).  

\begin{figure*}
\includegraphics[width = 7in]{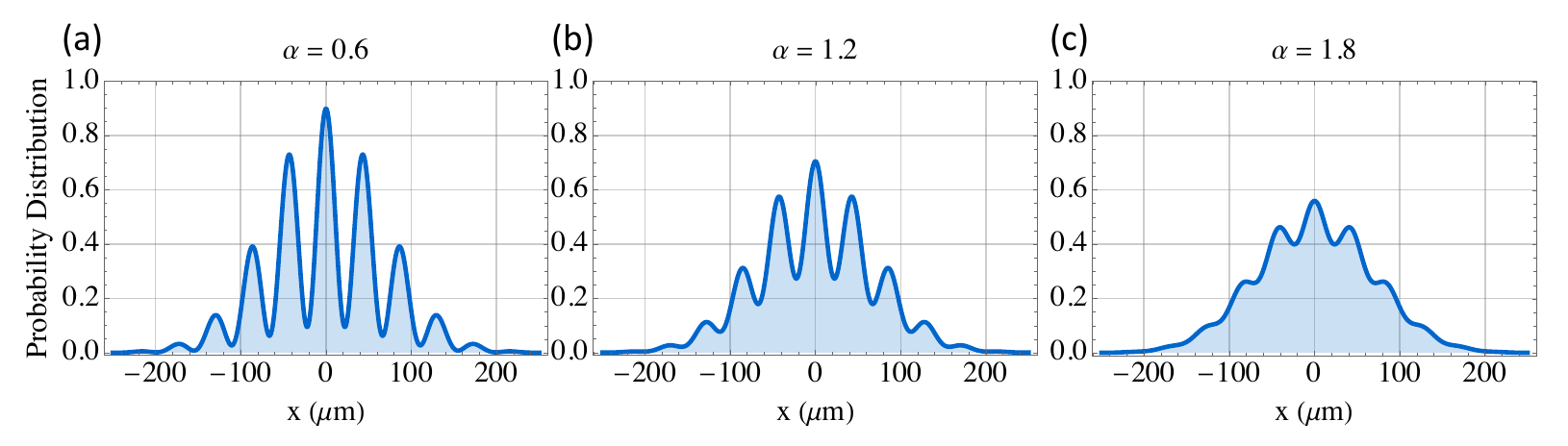}
\caption{\label{fig2} Interference pattern for electron double-slit experiment with monitoring.  Probability distribution (in arbitrary units) at time $t = 16.5 \ \mbox{ns}$ as a function of position $x$ for (a) interaction parameter $\alpha = 0.6$, recoil velocity $v_0 = 0.26 \ \mu \mbox{m/ns}$, and visibility $\mathcal{V} = 0.8$,  (b) interaction parameter $\alpha = 1.2$, recoil velocity $v_0 = 0.52 \ \mu \mbox{m/ns}$, and visibility $\mathcal{V} = 0.42$, and (c) interaction parameter $\alpha = 1.8$, recoil velocity $v_0 = 0.78 \ \mu \mbox{m/ns}$, and visibility $\mathcal{V} = 0.14$.  Other parameters include the slit width $d = 272 \ \mbox{nm}$ and the initial electron width $\delta = 20 \ \mbox{nm}$.}
\end{figure*}

Returning to the model of Sec.~II, we now consider the momentum exchange between the electron and proton.  First, an interaction timescale $\tau = 10 d/v \approx 190 \ \mbox{fs}$ is indeed much smaller than the characteristic spreading time $m \delta^2 /\hbar = 3.45\ \mbox{ps}$.  Second, the recoil velocity of the electron $v_0 \approx 260 \ \mbox{m/s}$ is much smaller than $v$.  This recoil leads to a small change of the overall interference pattern, as $v_0 T \approx 4.3 \ \mu\mbox{m}$ is small compared to $\Lambda \approx 44 \ \mu\mbox{m}$.   Finally, the visibility depends on $\alpha = 0.6$, $d$, and the width parameter $\Delta$ for the proton.  The experimental visibility of $\mathcal{V} \approx 0.8$ thus corresponds to $\Delta \approx 210 \ \mbox{nm}$.  This value of $\Delta$, being of the same order as $d$, is surely unrealistic, and violates the assumptions of the Coulomb scattering model \footnote{A more reasonable value for this thought experiment might be $\Delta = 100 \ \mbox{nm }$, which would yield visibilities in the range $0.65 < \mathcal{V} < 0.95$ for $\alpha$ between $0.6$ and $1.8$.}  Nevertheless, it is somewhat remarkable is that this simple model of a single-particle environment for the electron produces any reasonable results at all!

As noted above, the parameter $\alpha$ could be modified by changing the velocity of the electrons entering the double-slit.  This will change both the recoil velocity and the visibility in Eq.~(\ref{patterneq}).  Examples of such patterns, using the full expression for the probability distribution, are shown in Figs.~2 (b) and (c).  Larger values of  $\alpha$, corresponding to larger scattering, leads to smaller visibility $\mathcal{V}$ and reduced coherence.  

\section{Entanglement}

As mentioned in Sec.~I, the loss of coherence is associated with the entanglement between the proton and the electron.  In this section I will make this relationship quantitative.  The total wave function for the electron and the proton in Eq.~(\ref{psiout}) has the form
\begin{equation}
\frac{1}{\sqrt{2}} |\psi_1\rangle |\Phi_1\rangle + \frac{1}{\sqrt{2}} |\psi_2\rangle |\Phi_2\rangle,
\label{psidirac}
\end{equation}
with electron states $|\psi_1\rangle$ and $|\psi_2\rangle$ and proton states $|\Phi_1\rangle$ and $|\Phi_2\rangle$, with $\langle \psi_1 | \psi_2\rangle \approx 0$ when $\delta \ll d$.  When $\langle \Phi_1 | \Phi_2\rangle = \mathcal{V} \ne 1$, this is an entangled state.   Roughly speaking, this state can be interpreted as
\begin{align}
& \frac{1}{\sqrt{2}} |\text{``electron in slit 1''}\rangle |\text{``proton moving up''} \rangle \nonumber \\
&+ \frac{1}{\sqrt{2}} |\text{``electron in slit 2''}\rangle |\text{``proton moving down''} \rangle.
\end{align}

The standard way to characterize the entanglement of such a state is to consider the reduced density matrix of the electron $\rho$ \footnote{See Sec. 12.5 of Nielsen and Chuang (Ref. 32) or Sec. 8.4 of Barnett (Ref. 33).}.  This object provides the quantum statistical description of the electron.  For the entangled state of Eq.~(\ref{psidirac}), the density matrix takes the form
\begin{align}
\rho = & \frac{1}{2} \langle \Phi_1 | \Phi_1\rangle \,  |\psi_1\rangle \langle \psi_1|  + \frac{1}{2} \langle \Phi_2 | \Phi_2\rangle \,  |\psi_2\rangle \langle \psi_2|  \nonumber \\
& + \frac{1}{2} \langle \Phi_2 | \Phi_1 \rangle \, |\psi_1 \rangle \langle \psi_2| +  \frac{1}{2} \langle \Phi_1 | \Phi_2 \rangle \, |\psi_2 \rangle \langle \psi_1|.
\label{rho_electron}
\end{align}
Treating $ |\psi_1\rangle$ and $|\psi_2\rangle$ as the (approximately orthogonal) basis vectors in a two-dimensional space, this has the matrix form
\begin{equation}
\rho = \frac{1}{2} \left( \begin{array}{cc} 1 & \mathcal{V} \\  \mathcal{V} & 1 \end{array} \right).
\label{rho_matrix}
\end{equation}

The entanglement of this state is quantified by the von Neumann entropy \footnote{See Sec. 11.3 of Nielsen and Chuang (Ref. 32) or Sec. 8.1 of Barnett (Ref. 33).} of $\rho$:
\begin{equation}
S(\rho) = - \text{Tr} \left( \rho \log_2 \rho \right),
\end{equation}
where the logarithm of $\rho$ is evaluated in the basis in which $\rho$ is diagonal, {\em i.e.} the basis formed by the eigenvectors of $\rho$.  In this basis 
\begin{equation}
\rho_{\text{diag}} = \frac{1}{2} \left( \begin{array}{cc} 1 + \mathcal{V} & 0 \\ 0 & 1-\mathcal{V} \end{array} \right).
\end{equation}
Thus,
\begin{equation}
S(\rho) = H_2\left(\frac{1}{2} + \frac{1}{2} \mathcal{V}\right),
\label{vnentropy}
\end{equation}
where I have introduced the binary entropy function
\begin{equation}
H_2(x) = - x \log_2 x - (1-x) \log_2(1-x).
\end{equation}

The function $H_2(p)$ is the Shannon entropy \cite{[{}][{See Sec. 11.1 of Nielsen and Chuang (Ref. 32) or Ch. 1 of Barnett (Ref. 33).}]}, in bits, for a binary discrete random variable with probabilities $p$ and $1-p$.  This entropy characterizes the uncertainty associated with the random variable.  When the variable is maximally uncertain, $p=1/2$, and the Shannon entropy $H_2(1/2) = 1$.  Conversely, when $p=0$ or $p=1$, there is no uncertainty, and $H_2(0) = H_2(1) = 0$.  

The von Neumann entropy has a similar role, characterizing the minimal uncertainty associated with possible measurements made on $\rho$.  This uncertainty, in terms of the Shannon entropy, is minimized for measurements performed in the basis in which $\rho$ is diagonal.  The diagonal entries, which are the eigenvalues of $\rho$, are the probabilities for the minimum uncertainty measurement.  Then, Eq.~(\ref{vnentropy}) shows that the von Neumann entropy $S(\rho)$ equals the Shannon entropy for these probabilities.

The eigenvalues for the density matrix of Eq.~(\ref{rho_matrix}), when $\mathcal{V} = 1$ are $1$ and $0$.  In terms of the quantum state of Eq.~(\ref{psidirac}), when $\mathcal{V}=1$ the state is not entangled, and it is certain that the state of the electron is $(|\psi_1\rangle + |\psi_2\rangle)/\sqrt{2}$, which happens to be the eigenvector of $\rho$ with eigenvalue equal to $1$.  Thus, there is no uncertainty, and one obtains $S(\rho) = H_2(1) = 0$.  

When $\mathcal{V} = 0$, the eigenvalues of $\rho$ are both $1/2$, for which $S(\rho) = H_2(1/2) = 1$.  Here the state of the electron is maximally correlated with the state of the proton.  Thus, considering the electron on its own, its state is maximally uncertain, being either $|\psi_1\rangle$ or $|\psi_2\rangle$, each occurring with probability of $1/2$. This corresponds to the maximal possible uncertainty, and the entropy takes its maximal value.  Given that this is a binary set of possibilities, this corresponds to $1$ bit of information.  

\section{Information Gain}

The use of information and entropy to characterize the relationship between entanglement and the visibility of the interference pattern was introduced by Wootters and Zurek \cite{wootters1979complementarity} and recently discussed by Kincaid, McLelland, and Zwolak \cite{kincaid2016measurement}, the latter using a quantity known as the quantum mutual information.  This quantity arises in the field of quantum information \cite{nielsen2010quantum,barnett2009quantumbook}, which explores the ways in which quantum states can encode, transmit, and manipulate various types of information.  

Here I consider the classical information one can obtain about which path the electron takes through the double-slit---the ``which-way'' information. The relevant question is how much which-way information can be obtained from the proton about the potential path of the electron.  That is, we wish to determine which of the two possibilities (whether the electron passes through slit 1 or 2) has occurred, given some measurement of the proton.  Similar to the calculation of Eq.~(\ref{rho_electron}), the reduced density matrix for the proton alone is
\begin{equation}
\rho_{\text{proton}} =  \frac{1}{2} |\Phi_1\rangle \langle \Phi_1 | + \frac{1}{2}  |\Phi_2\rangle \langle \Phi_2 | ,
\label{rho_proton}
\end{equation}
and yet we know that $\rho_1 = |\Phi_1\rangle \langle \Phi_1|$ and $\rho_2 = |\Phi_2\rangle \langle \Phi_2|$ are correlated with the two states of the electron.  Thus, one might try to  determine through which slit the electron has passed by performing a measurement on the proton to distinguish between $\rho_1$ and $\rho_2$.  However, the fact that these proton states are not orthogonal means that this determination cannot be done exactly, and will involve some uncertainty.  

The initial uncertainty can be quantified in terms of the entropy and is simply $H_2(1/2) = 1$ bit, since there is no way to predict the state of the electron.  The final uncertainty, after some measurement on the proton, will be given by $H_2(p)$, where $p$ is the (post-measurement) probability for the path of the electron ({\it e.g.} for passing through slit 1).  The information {\em gained} from the measurement is the reduction in our uncertainty, or
\begin{equation}
I_{\text{gain}} = 1- H_2(p).
\end{equation}
In general, this information will depend on both the type of measurement performed and the particular outcome of that measurement.  To account for the latter, one should in fact average $I_{\text{gain}}$ over the different outcomes.  This average corresponds to the classical mutual information \footnote{See Sec. 11.2 of Nielsen and Chuang (Ref. 32) or Sec. 1.3 of Barnett (Ref. 33).} between the states of the electron and the measurements of the proton.  In this section I will discuss the various answers to the question of how much information can be gained; full calculations are presented in Appendix B, and the relationship with the quantum mutual information is discussed in Appendix C.

I will consider four different measurement scenarios.  The first is a direct measurement of the momentum, in the spirit of the original Bohr-Einstein debate.  This involves a continuous range of outcomes, and thus our average involves an integration over all possible vales of momentum.   The second, suggested by Maudlin, is to measure the deflection the proton up or down.  This provides a simpler binary outcome (``up'' or ``down''), at the cost of ignoring some information.  The third case, whose value of $I_{\text{gain}}$ was first obtained by Wootters and Zurek \cite{wootters1979complementarity}, corresponds to the optimal discrimination of non-orthogonal states $|\Phi_1\rangle$ and $|\Phi_2\rangle$.  As a fourth case, I consider the unambiguous state discrimination protocol recently discussed by Qureshi \cite{qureshi2016quantitative}.  Finally, these are all limited by the Holevo bound on the accessible information \footnote{See Sec. 12.1 of Nielsen and Chuang (Ref. 32).}, which in this case is precisely given by the von Neumann entropy calculated in Sec.~III.  The main results are summarized in Fig.~3---for all of the measurement methods, larger gains of information are associated with smaller visibility.  

\begin{figure}[t]
\includegraphics[width = 3.5in]{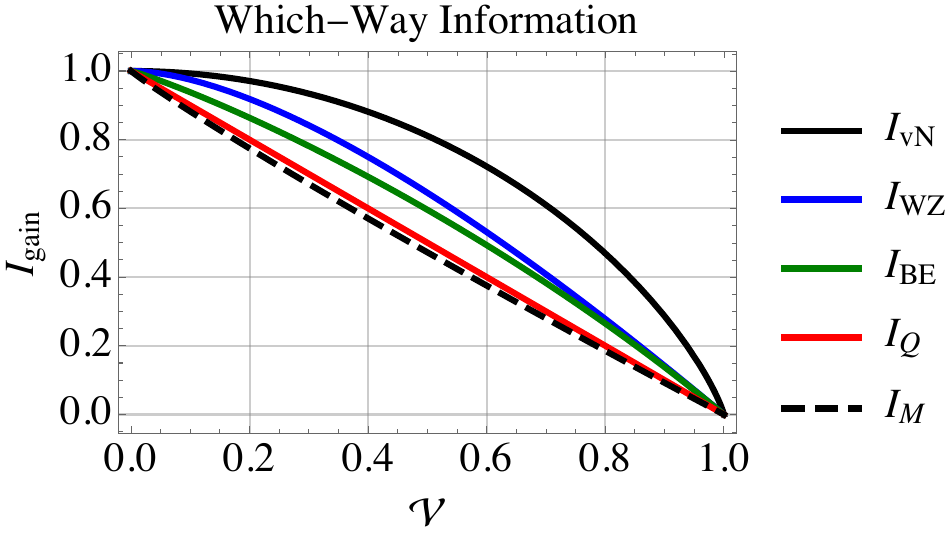}
\caption{\label{fig3} Information gain (in bits) as a function of visibility $\mathcal{V}$ for various measurement methods (see text).  From top to bottom: the upper bound $I_{\text{vN}}$ (black), the information gain from an optimal measurement $I_{\text{WZ}}$ (blue), the average information gain from a momentum measurement $I_{\text{BE} }$ (green), the average information gain from unambiguous state discrimination $I_{\text{Q}}$ (red), and the information gain from a binary (``up'' or ``down'') momentum measurement $I_{\text{M}}$ (black dashed).}
\end{figure}

\subsection{Full Momentum Measurement}

A momentum measurement is, in fact, the method originally discussed by Bohr and Einstein \cite{Bohr1949} and analyzed in \cite{wootters1979complementarity}.  This measurement yields a continuous set of outcomes.   For large positive momentum, one can be fairly certain that the electron has passed through slit 1 (and similarly for negative momentum and slit 2).  For values of the momentum near zero, however, one will remain uncertain.  

Using the momentum-space wave function for the proton, outcome $k$ occurs with probability density
\begin{equation}
\mathcal{P}(k) = \frac{1}{2} |\Phi_1(k)|^2 + \frac{1}{2} |\Phi_2(k)|^2,
\label{BEprobs}
\end{equation}
where $\Phi_1(k)$ and $\Phi_2(k)$ are given by Eqs.~(\ref{momentum1}) and (\ref{momentum2}).  Given this outcome, the conditional probability for the electron to pass through slit 1 is found by using Bayes' rule:
\begin{equation}
p(k) = \frac{1}{2} |\Phi_1(k)|^2 / \mathcal{P}(k).
\end{equation}
This is gives a post-measurement entropy $H_2\left( p(k) \right)$ that depends on the measurement outcome $k$.  The average gain of information is
\begin{equation}
I_{\text{BE}} = 1 - \int_{-\infty}^{+\infty} \mathcal{P}(k) H_2 \left( p(k) \right) dk.
\label{BEinfo}
\end{equation}

\subsection{Binary Momentum Measurement}

Maudlin's presentation \cite{maudlin2019philosophy} outfits the system with a detector that will register if the proton moves up or down.  These occur with probabilities
\begin{align}
p(\text{``up''}) &= \int_0^{\infty} \mathcal{P}(k) dk = \frac{1}{2}, \nonumber \\
p(\text{``down''}) &= \int_{-\infty}^{0} \mathcal{P}(k) dk = \frac{1}{2}. 
\end{align}
Given the outcome ``up'', the conditional probability for the electron to pass through slit 1 is found using Bayes' rule:
\begin{equation}
p(1| \text{``up''}) = \int_{0}^{+\infty} |\Phi_1(k)|^2 dk.
\end{equation}
Similarly, 
\begin{equation}
p(2| \text{``down''}) = \int_{-\infty}^{0} |\Phi_2(k)|^2 dk.
\end{equation}
These probabilities turn out to be identical, and the integral can be evaluated in terms of the error function:
\begin{align}
\frac{1}{\sqrt{\pi}} \int_{-u_0}^{+\infty} e^{-u^2} du & = \frac{1}{\sqrt{\pi}} \int_{-\infty}^{u_0} e^{-u^2} du  \nonumber \\
& = \frac{1}{2} \left( 1 + \frac{2}{\sqrt{\pi}} \int_{0}^{u_0} e^{-u^2} du \right) \nonumber \\
& = \frac{1}{2} \left[ 1 + \text{erf} (u_0) \right],
\end{align}
where I have set $u_0 = P \Delta / \hbar$.  Since these conditional probabilities are equal, the information gain does not depend on the measurement outcome.  One thus finds, using the fact that $\mathcal{V} = e^{-u_0^2}$, that the information gain is
\begin{equation}
I_{\text{M}} = 1 - H_2\left( 1 + \text{erf}(\sqrt{-\ln \mathcal{V}}) \right).
\label{Minfo}
\end{equation}

\subsection{Minimum-Error State Discrimination}

To obtain the maximum information gain, for general $\mathcal{V}$, one must use the measurement that best distinguishes between $|\Phi_1\rangle$ and $|\Phi_2\rangle$; this is known as minimum-error state discrimination \cite{[{}][{, See also Sec. 4.4 of Barnett (Ref. 33).}]barnett2009quantum}. The optimal measurement (originally due to Helstrom) uses the basis vectors $|e_1\rangle$ and $|e_2\rangle$ chosen so that
\begin{align}
|\Phi_1\rangle &= \cos (\theta/2) |e_1\rangle + \sin(\theta/2) |e_2\rangle \nonumber \\
|\Phi_2\rangle &= \sin (\theta/2) |e_1\rangle + \cos (\theta/2) |e_2\rangle,
\label{WZstates}
\end{align} 
where $\langle \Phi_1 | \Phi_2 \rangle = \sin \theta = \mathcal{V}$.  

Outcomes $|e_1\rangle$ and $|e_2\rangle$ both occur with total probability $1/2$.  The conditional probability for the electron to pass through slit 1 given outcome $|e_1\rangle$ (or slit 2 given outcome $|e_2\rangle$) is given by Bayes' rule:
\begin{align}
p &= |\langle e_1 | \Phi_1 \rangle|^2 = |\langle e_2 | \Phi_2 \rangle|^2 \nonumber \\
&= \frac{1}{2} (1+ \cos \theta ) \nonumber \\
&= \frac{1}{2} \left( 1+ \sqrt{1 - \mathcal{V}^2} \right).
\end{align}
Once again, as these conditional probabilities are equal, the information gain does not depend on the measurement outcome.  The information gained is thus
\begin{equation}
I_{\text{WZ}} = 1 - H_2\left(\frac{1}{2} + \frac{1}{2} \sqrt{1 - \mathcal{V}^2}\right).
\end{equation}
This result was first obtained by Wootters and Zurek \cite{wootters1979complementarity}, who proved that $I_{\text{gain}} \le I_{\text{WZ}}$.  The general quantum information problem, involving what is called the accessible information \cite{nielsen2010quantum}, was analyzed in \cite{fuchs1994ensemble} and \cite{levitin1995optimal}. 

\subsection{Unambiguous State Discrimination}

Another measurement method, known as unambiguous state discrimination \cite{barnett2009quantumbook,barnett2009quantum}, was recently discussed by Qureshi \cite{qureshi2016quantitative}.  This method uses an interaction with an auxiliary system designed so that a measurement of that system indicates whether or not the determination of the electron's position will be successful.  

For a successful determination, which occurs with probability $1-\mathcal{V}$, one obtains a result with certainty, so that $p=1$.  For an unsuccessful determination, however, one remains ignorant of the electron's position, so that $p=1/2$.  The average gain of information is thus 
\begin{equation}
I_{Q} = 1 - \left[ (1 - \mathcal{V}) H_2(1) + \mathcal{V} H_2(1/2) \right] = 1 - \mathcal{V}.
\end{equation}

\subsection{Holevo bound}

The Holevo theorem \cite{nielsen2010quantum,fuchs1994ensemble} bounds the accessible information associated with the density matrix of Eq.~(\ref{rho_proton}) by
\begin{equation}
I_{\text{gain}} \le S(\rho_{\text{proton}}) - \frac{1}{2} S(\rho_1) - \frac{1}{2} S(\rho_2),
\label{holevo1}
\end{equation}
where $S(\rho_j)$ is the von Neumann entropy of $\rho_j$.  However (and shown in Appendix B), $\rho_1$ and $\rho_2$ are both pure states (with eigenvalues 0 and 1), and thus have zero entropy, while the density matrix $\rho_{\text{proton}}$ has the same eigenvalues as those calculated in Sec.~IV.  Thus, the right-hand-side of Eq.~(\ref{holevo1}) reduces to the von Neumann entropy calculated in Eq.~(\ref{vnentropy}):
\begin{equation}
I_{\text{vN}} =  H_2\left(\frac{1}{2} + \frac{1}{2} \mathcal{V} \right).
\end{equation}

\subsection{Comparison of Information Gained}

The information gain for these different scenarios is plotted as a function of the visibility $\mathcal{V}$ in Fig.~3, where I have numerically integrated Eq.~(\ref{BEinfo}).   As can be seen, the amount of information gained depends on the measurement, and is ultimately bounded by the measure of entanglement, satisfying the inequalites
\begin{equation}
I_{\text{M}} \le I_{Q} \le I_{\text{BE}} \le I_{\text{WZ}} \le I_{\text{vN}}.
\end{equation}
Note, however, that it is not possible to attain $I_{\text{vN}}$ using just one measurement when $0< \mathcal{V} < 1$ \cite{fuchs2002just}, thus one generally has $I_{\text{gain}} \le  I_{\text{WZ}}$.  

\section{Conclusion}

In this paper I have explored the decoherence model for the double-slit experiment introduced by Maudlin \cite{maudlin2019philosophy}.   This simple model shows that decoherence does not require significant calculations or a complicated model of the environment---even a single photon (or proton) is sufficient to destroy the interference pattern for the electron.   It also highlights the role of entanglement in decoherence mechanisms.  The quantitative analysis presented here is suitable for an undergraduate course in quantum mechanics, requiring only Gaussian wave packets and integrals.  Thus, it could be a useful starting point for students to learn about decoherence \cite{zurek2003decoherence, schlosshauer2007decoherence,schlosshauer2019quantum}.  

The presentation here could also be used to introduce ideas from quantum information theory \cite{nielsen2010quantum,barnett2009quantumbook}, such as the Shannon and von Neumann entropies, the mutual information, and the Holevo bound.  In particular, the discussion of the different measurement scenarios in Sec.~V (and in Appendix B) resulting in the different information curves in Fig.~3 provides a simple yet sufficiently rich example to introduce these topics.  Furthermore, comparison of the results presented here with the work of Kincaid, McLelland, and Zwolak \cite{kincaid2016measurement} and Qureshi \cite{qureshi2016quantitative} would give students a common starting point to explore a broad set of ideas in quantum information.

At a fundamental level, one may worry that the semiclassical model of momentum exchange between the electron and the proton given here is too simple.  This can be improved upon by using a more involved (yet still one-dimensional) approach described in Appendix A which preserves the Gaussian form of the wave function (albeit with a more complicated time-evolution).  One could possibly extend this through numerical simulations, or by extending the one-dimensional treatment to a fully three-dimensional quantum calculation using scattering theory \footnote{See Ch. 21 of Bohm (Ref. 11).}.  These exercises are left for ambitious readers.

Regardless of the context or application, it is surprising to see just how much information about the mystery of quantum mechanics can be gained by analyzing the double-slit experiment---the ``experiment with the two holes.''  

\begin{acknowledgments}
I sincerely thank Keith McPartland, Isaac Wilkins, and Bill Wootters for many engaging, fruitful, and educational discussions on decoherence.  I also thank the late David Park for providing a prescient comment.  Finally, I thank the three anonymous referees for their comments on an earlier version of this manuscript.
\end{acknowledgments}

\appendix

\section{Interference Pattern Calculations}

The interference pattern in Sec.~III requires one to find the time-dependent wave function  $\Psi(x,X,t)$ given the initial condition $\Psi(x,X,t=0) = \Psi_{\text{out}}(x,X)$ using Eq.~(\ref{psiout}).  This can be performed by using the classic result for a Gaussian wave packet \footnote{See Ch. 3 Secs. 5 and 13 of Bohm (Ref. 11)}  starting with width $\delta$, initial position $x_0$, and initial momentum $p_0$:
\begin{widetext}
\begin{equation}
\psi_{\text{G}}(x,t; x_0,p_0) =\pi^{-1/4} \sqrt{\frac{\delta}{\delta^2 + i \hbar t/m}}  \exp\left[ - \frac{1}{2} \frac{ (x-x_0 - p_0 t/m)^2}{\delta^2 + i \hbar t/m} +  i p_0 x / \hbar - i \frac{p_0^2 t}{2 m \hbar} \right].
\label{gaussian}
\end{equation}
To use this result, observe that for the wave functions of Eqs.~(\ref{psi1})-(\ref{psip})
\begin{equation}
\begin{array}{lcl}
\psi_1(x) e^{-i P x/\hbar} &=& \psi_{\text{G}}(x,0; +d/2,-P), \\
\psi_2(x) e^{+i P x/\hbar} &=& \psi_{\text{G}}(x,0; -d/2,+P), \\ 
\psi_{\text{p}}(X) e^{\pm i P X /\hbar} &=& \tilde{\psi}_{\text{G}}(X,0; 0,\pm P),
 \end{array}
 \end{equation}
where $\tilde{\psi}_{\text{G}}(X,t;x_0,p_0)$ has the same form as Eq.~(\ref{gaussian}) with $\delta \to \Delta$ and $m \to M$.    Using this result, the total wave function at time $t$ is
\begin{equation}
\Psi(x,X,t) = \frac{N}{\sqrt{2}} \psi_{\text{G}}(x,t;+d/2, -P) \ \tilde{\psi}_{\text{G}}(X,t; 0, +P) + \frac{N}{\sqrt{2}} \psi_{\text{G}}(x,t;-d/2,+P) \ \tilde{\psi}_{\text{G}}(X,t;0,-P),
\label{psitotal}
\end{equation}
\end{widetext}
where the normalization factor can be calculated at time $t=0$ (and will remain constant in time) as
\begin{equation}
N = \left( 1 + e^{-(d/2)^2/\delta^2}e^{ -P^2 \delta^2/\hbar^2}e^{ -P^2 \Delta^2/\hbar^2} \right)^{-1/2}.
\end{equation}
Note also that the visibility $\mathcal{V}$ can also be calculated at time $t=0$ (and will also remain constant in time).

Note the wave function can be represented in a mixed form
\begin{widetext}
\begin{equation}
\Psi(x,k,t) = \frac{N}{\sqrt{2}} e^{- \hbar k^2 t/2M} \left[ \psi_{\text{G}}(x,t; +d/2,-P) \ \Phi_1(k) + \psi_{\text{G}}(x,t;-d/2,+P) \ \Phi_2(k) \right],
\end{equation}
\end{widetext}
where $\Phi_1(k)$ and $\Phi_2(k)$ are the momentum space wave functions for the proton moving ``up'' or ``down'' given in Eqs.~(\ref{momentum1}) and (\ref{momentum2}), with mean velocity $\pm P/M$.  This allows for an interpretation of the loss of coherence due to the conditional displacement of the proton in momentum space [30], which is illustrated in Fig. 4.

\begin{figure*}
\includegraphics[width = 7in]{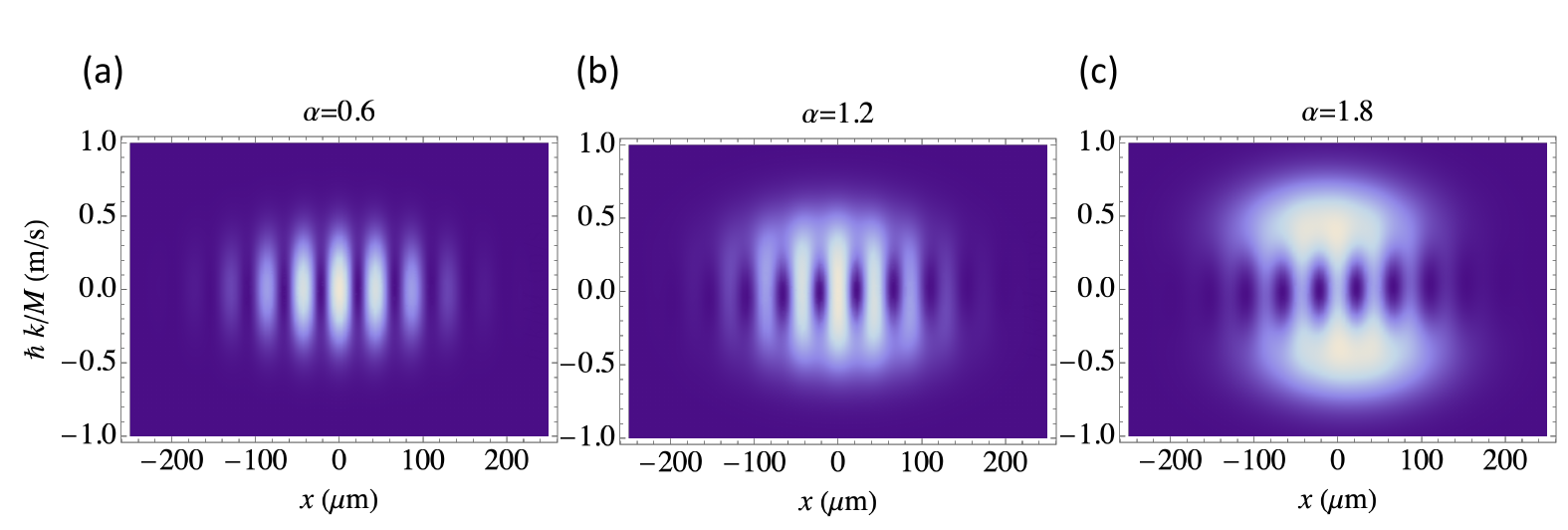}
\caption{\label{fig4} Wave function for electron double-slit experiment with monitoring.  Probability distribution (in arbitrary units) at time $t = 16.5 \ \mbox{ns}$ as a function of electron position $x$ and proton velocity $\hbar k/M$ for  (a) interaction parameter $\alpha = 0.6$, electron recoil velocity $v_0 = 0.26 \ \mu \mbox{m/ns}$, proton velocity $P/M = 0.14 \ \mbox{m/s}$, and visibility $\mathcal{V} = 0.8$,  (b) interaction parameter $\alpha = 1.2$, electron recoil velocity $v_0 = 0.52 \ \mu \mbox{m/ns}$, proton velocity $P/M = 0.28 \ \mbox{m/s}$, and visibility $\mathcal{V} = 0.42$, and (c) interaction parameter $\alpha = 1.8$, recoil velocity $v_0 = 0.78 \ \mu \mbox{m/ns}$, proton velocity $P/M = 0.42 \ \mbox{m/s}$, and visibility $\mathcal{V} = 0.14$.  Other parameters include the slit width $d = 272 \ \mbox{nm}$ and the initial electron and proton widths $\delta = 20 \ \mbox{nm}$ and $\Delta = 210 \ \mbox{nm}$.}
\end{figure*}

A more challenging calculation (but still restricted to one spatial dimension) would involve going beyond a simple momentum exchange and treating the interaction as an impulsive Coulomb interaction
\begin{equation}
\Psi_{\text{out}}(x,X) = \exp \left(- \frac{i}{\hbar} \int_{-\tau/2}^{\tau/2} V(x,X,t) dt \right) \Psi_{\text{in}}(x,X),
\end{equation}
where
\begin{equation}
V(x,X,t) = - \frac{e^2}{4 \pi \epsilon_0} \frac{1}{\sqrt{(X-x)^2 +  (v t)^2}}.
\end{equation}
Assuming that $|x-X| \approx d/2$, one can then expand $V(x,X,t)$ as a Taylor series (in $(x-X)^2 - d^2/4$) to find
\begin{equation}
V(x,X,t) \approx V_0(t) + \frac{e^2}{8 \pi \varepsilon_0} \frac{(x-X)^2}{ \left[d^2/4 + (v t)^2 \right]^{3/2} },
\end{equation}
where $V_0(t)$ does not depend on $x$ or $X$.  Thus, up to an overall phase, 
\begin{equation}
\Psi_{\text{out}}(x,X) \approx e^{-i a (x-X)^2} \Psi_{\text{in}}(x,X),
\end{equation}
where
\begin{align}
a &= \frac{e^2}{8 \pi \varepsilon_0 \hbar} \int_{-\tau/2}^{\tau/2} \frac{dt}{\left[ d^2/4 + (v t)^2 \right]^{3/2}} \nonumber \\
& = \frac{e^2}{8 \pi \varepsilon_0 \hbar} \left( \frac{8}{v d^2} \right) \frac{( v \tau / d)}{\sqrt{1 + (v \tau/d)^2}} \nonumber \\
& \approx \frac{e^2}{\pi \varepsilon_0 v \hbar} \frac{1}{d^2} = \alpha / d^2.
\end{align}
The subsequent time-evolution can be calculated exactly by a patient application of Gaussian integrals.  The results are largely similar to that described above, albeit with some different coefficients in the resulting interference pattern.  For example, the visibility is given by
\begin{equation}
\mathcal{V} = e^{-\alpha^2 \Delta^2/(d^2 + 4 \alpha^2 \delta^2 \Delta^2/d^2)}.
\end{equation} 

\section{Classical Mutual Information Calculations}

In this Appendix I provide full calculations for the results of Sec.~V, using notation and terminology from quantum information theory \cite{nielsen2010quantum,barnett2009quantumbook}.   What I called $I_{\text{gain}}$ is actually the classical mutual information between two random variables.  The first, denoted by $X$, corresponds to the events ``electron in slit 1'' ($x=1$) and ``electron in slit 2'' ($x=2$), each occurring with probability $p_{x=1} = p_{x=2} = 1/2$.  The second, denoted by $Y$, corresponds to the measurement outcomes and associated probabilities when the proton is subject to one of the measurements described in Sec.~V.  The mutual information between $X$ and $Y$ is quantified by
\begin{equation}
I(X:Y) = H(X) + H(Y) - H(X,Y),
\end{equation}
where the quantities $H(X), H(Y), \ \mbox{and} \ H(X,Y)$ are the Shannon entropies for the random variables $X$, $Y$, and their combination $(X,Y)$.  Given the joint probability distribution $p_{x,y}$ for $(X,Y)$, the total Shannon entropy is
\begin{equation}
H(X,Y) = - \sum_{x,y} p_{x,y} \log_2 p_{x,y}.
\end{equation}
The marginal probability distributions for $X$ and $Y$ are 
\begin{align}
p_{x} = & \sum_{y} p_{x,y}, \nonumber \\
p_{y} = & \sum_{x} p_{x,y},
\end{align}
in terms of which
\begin{align}
H(X) = - \sum_{x} p_x \log_2 p_x, \nonumber \\
H(Y) = - \sum_{y} p_y \log_2 p_y.
\end{align}
As information theory is not traditionally covered in the undergraduate physics curriculum, I will provide detailed calculations of these quantities for each of the measurement methods discussed in Sec.~V.  The Holevo bound on the mutual information will also be calculated.  

Note that in Sec.~V I calculated the mutual information in the alternative form
\begin{equation}
I(X:Y) = H(X) - H(X|Y),
\end{equation}
where the conditional entropy is
\begin{align}
H(X|Y) &= - \sum_{y} p_y H(X | Y = y) \nonumber \\
& = -\sum_{x,y} p_y p(x|y) \log_2 p(x|y)
\end{align}
with the conditional probabilities
\begin{equation}
p(x|y) = p_{x,y}/p_y.
\end{equation}
One can show that $H(X|Y) = H(X,Y) - H(Y)$.  

\subsection{Calculation of $I_{\text{BE}}$}

For a full momentum measurement there is a continuous set of measurement outcomes, with $\mathcal{P}_x(k)$ taking the place of $p_{x,y}$.  The joint probabilities are given by Eq.~(\ref{BEprobs}) and Eqs.(\ref{momentum1}-\ref{momentum2}):
\begin{align}
\mathcal{P}_1(k) &= \frac{1}{2} \pi^{-1/2} \Delta \exp \left[ -\Delta^2 (k - P/\hbar)^2 \right], \nonumber \\
\mathcal{P}_2(k) &= \frac{1}{2} \pi^{-1/2} \Delta \exp \left[ -\Delta^2 (k + P/\hbar)^2 \right].
\label{BEprob2}
\end{align}
The marginal probabilities are
\begin{align}
p_{x=1} &= \int_{-\infty}^{\infty} p_1(k) dk = \frac{1}{2}, \nonumber \\
p_{x=2} &= \int_{-\infty}^{\infty} p_2(k) dk = \frac{1}{2}, \nonumber \\
\mathcal{P}(k) &= \mathcal{P}_1(k) + \mathcal{P}_2(k).
\end{align}
Using these probabilities, the entropies are $H(X)=1$, 
\begin{equation}
H(Y) = - \int_{-\infty}^{\infty} \mathcal{P}(k) \log_2 \mathcal{P}(k) dk, 
\end{equation}
and
\begin{equation}
H(X,Y) = - \int_{-\infty}^{\infty} \left[ \mathcal{P}_1(k) \log_2 \mathcal{P}_1(k) + \mathcal{P}_2(k) \log_2 \mathcal{P}_2(k) \right] dk, 
\end{equation}
Given these entropies, the mutual information $I(X:Y)$ equals the expression for $I_{\text{BE}}$ in Eq.~(\ref{BEinfo}). 

\subsection{Calculation of $I_{\text{M}}$}

For the binary momentum measurement of Maudlin \cite{maudlin2019philosophy} there are two outcomes, corresponding to ``proton moving up'' ($y=\text{``up''}$) and ``proton moving down'' ($y=\text{``down''}$).  The joint probabilities, using Eq.~(\ref{BEprob2}), are
\begin{align}
p_{1,\text{up}} &= \int_{0}^{\infty} \mathcal{P}_1(k) dk = \frac{1}{4} [1 + \text{erf}(u_0)], \nonumber \\
p_{1,\text{down}} &= \int_{-\infty}^{0} \mathcal{P}_1(k) dk = \frac{1}{4} [1 - \text{erf}(u_0)], \nonumber \\
p_{2,\text{up}} &= \int_{0}^{\infty} \mathcal{P}_2(k) dk = \frac{1}{4} [1- \text{erf}(u_0)], \nonumber \\
p_{2,\text{down}} &= \int_{-\infty}^{0} \mathcal{P}_2(k) dk = \frac{1}{4} [1 + \text{erf}(u_0)],
\end{align}
where $u_0 = P \Delta /\hbar$.  The marginal probabilities are
\begin{align}
p_{x=1} &= p_{1,\text{up}} + p_{1,\text{down}} = \frac{1}{2},  \nonumber \\
p_{x=2} &= p_{2,\text{up}} + p_{2,\text{down}} = \frac{1}{2},  \nonumber \\
p_{\text{up}} &= p_{1,\text{up}} + p_{2,\text{up}}= \frac{1}{2},  \nonumber \\
p_{\text{down}} &= p_{1,\text{down}} + p_{2,\text{down}} = \frac{1}{2}.
\end{align}
The resulting entropies are $H(X) = H(Y) = 1$ and
\begin{equation}
H(X,Y) = 1 + H_2\left( \frac{1}{2} + \frac{1}{2} \text{erf}(u_0) \right).
\end{equation}
The mutual information for this measurement is
\begin{equation}
I(X:Y) = 1 - H_2\left( \frac{1}{2} + \frac{1}{2} \text{erf}(u_0) \right),
\end{equation}
which agrees with the expression for $I_{\text{M}}$ in Eq.~(\ref{Minfo}).

\subsection{Calculation of $I_{\text{WZ}}$}

For the optimal state discrimination approach \cite{barnett2009quantumbook, barnett2009quantum}, the measurement outcomes are $\{ |e_1\rangle , |e_2\rangle \}$, in terms of which with the states $\{ |\Phi_1\rangle , |\Phi_2\rangle \}$ are related by Eq.~(\ref{WZstates}).  The joint probabilities $p_{x,y} = \frac{1}{2} |\langle e_y | \Phi_x \rangle|^2$ are found to be
\begin{align}
p_{1,1} = p_{2,2} &= \frac{1}{2} \cos^2(\theta/2) = \frac{1}{4} (1+ \cos \theta), \nonumber \\
p_{1,2} = p_{2,1} &= \frac{1}{2} \sin^2(\theta/2) = \frac{1}{4} (1-\cos \theta).
\end{align}
The marginal probabilities are calculated as
\begin{align}
p_{x=1} &= p_{1,1} + p_{1,2} = \frac{1}{2}, \nonumber \\
p_{x=2} &= p_{2,1} + p_{2,2} = \frac{1}{2}, \nonumber \\
p_{y=1} &= p_{1,1} + p_{2,1} = \frac{1}{2}, \nonumber \\
p_{y=2} &= p_{1,2} + p_{2,2} = \frac{1}{2}.
\end{align}
Thus, $H(X) = H(Y) = 1$, and 
\begin{align}
H(X,Y) =&- \frac{1}{2} (1 + \cos \theta) \log_2 \left[ \frac{1}{4} (1 + \cos \theta) \right] \nonumber \\
& - \frac{1}{2} (1 - \cos \theta) \log_2 \left[ \frac{1}{4} (1 - \cos \theta) \right] \nonumber \\
=& 1 + H_2\left( \frac{1}{2} + \frac{1}{2} \cos \theta \right).
\end{align}
Using $\cos \theta = \sqrt{1 - \sin^2 \theta} = \sqrt{1 - \mathcal{V}^2}$, the mutual information evaluates to
\begin{equation}
I(X:Y) = I_{\text{WZ}} = 1 - H_2 \left(\frac{1}{2} + \frac{1}{2} \sqrt{1 - \mathcal{V}^2}\right).
\end{equation}

\subsection{Calculation of $I_{\text{Q}}$}

For the unambiguous state discrimination approach \cite{qureshi2016quantitative, barnett2009quantumbook,barnett2009quantum}, there are actually three measurement outcomes.  The two successful outcomes are labelled by $y=1$ and $y=2$, while the third outcome $y=3$ corresponds to failure.  The nonzero joint probabilities $p_{x,y}$ are found to be 
\begin{align}
p_{1,1} = p_{2,2} &= \frac{1}{2} (1 - \mathcal{V}), \nonumber \\
p_{1,3} = p_{2,3} &= \frac{1}{2} \mathcal{V}.
\end{align}
The marginal probabilities are
\begin{align}
p_{x=1} &= p_{1,1} + p_{1,2} + p_{1,3} = \frac{1}{2}, \nonumber \\
p_{x=2} &= p_{2,1} + p_{2,2} + p_{2,3} = \frac{1}{2}, \nonumber \\
p_{y=1} &= p_{1,1} + p_{2,1} = \frac{1}{2}(1-\mathcal{V}), \nonumber \\
p_{y=2} &= p_{1,2} + p_{2,2} = \frac{1}{2}(1-\mathcal{V}), \nonumber \\
p_{y=3} &= p_{1,3} + p_{2,3} = \mathcal{V}.
\end{align}
Using these probabilities, the entropies are 
\begin{align}
H(X) &= 1, \nonumber \\
H(Y) &= (1-\mathcal{V}) + H_2(1-\mathcal{V}), \nonumber \\
H(X,Y) &= 1 + H_2(1-\mathcal{V}),
\end{align}
and thus the mutual information is
\begin{equation}
I(X:Y) = I_{\text{Q}} = 1 - \mathcal{V}.
\end{equation}

\subsection{Calculation of the Holevo Bound}

Finally, the expression for the Holevo bound,
\begin{equation}
I(X:Y) \le S(\rho_\text{proton}) - \frac{1}{2} S(\rho_1) - \frac{1}{2} S(\rho_2),
\end{equation}
involves the von Neumann entropies of the density matrices
\begin{align}
\rho_1 = |\Phi_1\rangle \langle \Phi_1|, \nonumber \\
\rho_2 = |\Phi_2\rangle \langle \Phi_2|, \nonumber \\
\rho_{\text{proton}} =  \frac{1}{2} \rho_1+ \frac{1}{2} \rho_2.
\end{align}
These matrices can be expressed in terms of the basis states $|e_1\rangle$ and $|e_2\rangle$ using Eq.~(\ref{WZstates}), which yields
\begin{equation}
\rho_1 = \frac{1}{2} \left( \begin{array}{cc} 1 + \cos \theta & \sin \theta \\ \sin \theta & 1 - \cos \theta \end{array} \right),
\end{equation}
\begin{equation}
\rho_2 = \frac{1}{2} \left( \begin{array}{cc} 1 - \cos \theta & \sin \theta \\ \sin \theta & 1 + \cos \theta \end{array} \right),
\end{equation}
and
\begin{equation}
\rho_{\text{proton}} = \frac{1}{2} \left( \begin{array}{cc} 1 & \sin \theta \\ \sin \theta & 1 \end{array} \right).
\end{equation}
The corresponding eigenvalues are $\{1,0\}$, $\{1,0\}$, and $\frac{1}{2} (1 \pm \sin \theta)$, respectively.  Using $\sin \theta = \mathcal{V}$, we obtain
\begin{align}
S(\rho_1) &= H(1) = 0, \nonumber \\
S(\rho_2) &= H(1) = 0, \nonumber \\
S(\rho_{\text{proton}}) &= H\left(\frac{1}{2} + \frac{1}{2} \mathcal{V}\right),
\end{align}
so that
\begin{equation}
I(X:Y) \le H\left(\frac{1}{2} + \frac{1}{2} \mathcal{V}\right) = I_{\text{vN}}.
\end{equation}

\section{Quantum Mutual Information}
This classical mutual information analysis is appropriate if the electron has, in fact, been measured to be in one of the states $|\psi_1\rangle$ or $|\psi_2\rangle$, and not allowed to interfere.  This would be the case if the electron was coupled to both a perfect detector (with orthogonal states $|d_1\rangle$ and $|d_2\rangle$) and the proton (with nonorthogonal states $|\Phi_1\rangle$ and $|\Phi_2\rangle$).  The appropriate quantum state for this electron-proton-detector system is
\begin{equation} 
\frac{1}{\sqrt{2}} |\psi_1\rangle |\Phi_1\rangle |d_1\rangle + \frac{1}{\sqrt{2}} |\psi_2\rangle |\Phi_2\rangle |d_2\rangle.
\end{equation}
The classical mutual information is then associated with how much information one can gain about the electron by measuring the proton.  The corresponding quantum mutual information \cite{kincaid2016measurement} is
\begin{equation}
I_{\text{quantum}} = S(\rho_{\text{electron}}) + S(\rho_{\text{proton}}) - S(\rho_{\text{elec-prot}}),
\end{equation}
where the right-hand-side of this equation involves the von Neumann entropies of the density matrices
\begin{align}
\rho_{\text{electron}} = \frac{1}{2} |\psi_1\rangle \langle \psi_1| + \frac{1}{2} |\psi_2\rangle \langle \psi_2|, \nonumber \\
\rho_{\text{proton}} = \frac{1}{2} |\Phi_1\rangle \langle \Phi_1| + \frac{1}{2} |\Phi_2\rangle \langle \Phi_2|,
\end{align}
and
\begin{equation}
\rho_{\text{elec-prot}} = \frac{1}{2} |\psi_1\rangle \langle \psi_1| \otimes |\Phi_1 \rangle \langle \Phi_1| + \frac{1}{2} |\psi_2\rangle \langle \psi_2| \otimes |\Phi_2\rangle \langle \Phi_2|.
\end{equation}
Note that $\rho_{\text{electron}}$ differs from the density matrix in Sec.~IV because of the orthogonality of the detector states.  Since $S(\rho_{\text{electron}}) = S(\rho_{\text{elec-prot}}) = 1$, we have that $I_{\text{quantum}} = S(\rho_{\text{proton}}) = I_{\text{vN}}$, which coincides with the Holevo bound.  

\section{Author Declarations}

The authors have no conflicts to disclose.

\bibliography{double_slit}

\end{document}